\documentclass[print,pra,preprintnumbers,twocolumn]{revtex4}
\usepackage{amssymb}
\usepackage{dcolumn}
\usepackage{amsmath}
\usepackage{bm}
\usepackage{epsfig}
\usepackage{epstopdf}
\usepackage{graphicx}
\usepackage{ulem}
\usepackage[utf8]{inputenc}
\usepackage{amsmath}
\usepackage{amsthm}
\usepackage{bm}
\usepackage{color}

\begin{document}

\title{Beyond the Carnot Limit in the Internal Cycles of a Quantum Heat Engine under Finite Heat Reservoirs}
\author{L.-L.Yan$^{1,2}$}
\email{llyan@zzu.edu.cn}
\author{M.-R. Yun$^{1}$}
\author{M. Li$^{3}$}
\author{S.-L. Su$^{1,2}$}
\email{slsu@zzu.edu.cn}
\author{K.-F. Cui$^{1}$}
\author{Gang Chen$^{1,2}$}
\email{chengang971@163.com}
\author{M. Feng$^{4,5}$}
\email{mangfeng@wipm.ac.cn}

\affiliation{$^{1}$ Key Laboratory of Materials Physics, Ministry of Education, School of Physics and Laboratory of Zhongyuan Light, Zhengzhou University, Zhengzhou 450001, China\\
$^{2}$ Institute of Quantum Materials and Physics, Henan Academy of Sciences, Zhengzhou 450046, China \\
$^{3}$ School of Telecommunications and Intelligent Manufacturing, Sias University, Xinzheng,451150, China \\
$^{4}$ State Key Laboratory of Magnetic Resonance and Atomic and Molecular Physics,
Wuhan Institute of Physics and Mathematics, Innovation Academy of Precision Measurement Science and Technology, Chinese Academy of Sciences, Wuhan, 430071, China \\
$^{5}$ Research Center for Quantum Precision Measurement, Guangzhou Institute of Industry Technology, Guangzhou 511458, China }

\begin{abstract}
We investigate, in an analytical fashion, quantum Carnot cycles of a microscopic heat engine coupled to two finite heat reservoirs, whose internal cycles could own higher efficiency than the standard Carnot limit without consuming extra quantum resources, e.g., coherence or squeezing properties. The engine runs time-dependently, involving both the internal and external cycles to collaboratively accomplish a complete Carnot cycle, and the efficiency of the engine depends on the reservoirs' heat capacities and the working substance. Our analytical results of the maximum efficiency and the maximum power output clarify the mechanism behind the high performance of the microscopic engines, displaying the key roles played by the finite-sized heat reservoirs. Our proposal is generally valid for any microscopic thermodynamic system and fully feasible under current laboratory conditions.
\end{abstract}

\maketitle

Heat engines, converting heat into mechanical work, have played an essential role in exploring the fundamental nature and practical application of thermodynamics \cite{Carnot,Cengel}. With emergent techniques decreasing the length and energy scales, studies of the heat engines have been moving toward the microscopic region, even to quantum regime \cite{PRL-2-262,JPA-12-L103,JCP-80-1625,PRL-88-050602,PRE-76-031105,EPL-81-20003,PRX-5-031044,PRL-124-040602}. Along with theoretical developments, experimental efforts have been devoted to the engines in different physical systems from colloidal particles \cite{NP-8-143,PRL-114-120601,NP-12-67,NP-12-1134} to engineered quantum systems \cite{Science-352-325,PRAplied-6-054014,PRL-112-150601,PRL-122-110601,NP-14-911}, with some peculiar designs by consuming extra resources and performing customized cycles, e.g., quantum heat engine cycles \cite{PRL-97-180402,RMP-81-1,CPL-35-040301,PRL-123-250606,PRL-124-100603,PRL-106-070401,PNAS-111-13786,PRL-120-100601} and critical heat engines working on the exceptional conditions \cite{NC-7-11895,PRE-96-022143,NPJ-5-88}.

As a vital property of the engines, the Carnot limit \cite{Carnot} sets a standard reference bound on the efficiency of conventional heat engines due to the restriction
of the thermodynamic laws. Devoting to the Carnot limit in quantum regime, a seminal work \cite{Science-299-862} has shown the possibility to enhance the engine's efficiency beyond the standard Carnot limit by quantum resource of coherence, and then along this way, the standard Carnot limit was broken by other approaches, e.g., coupling the engine to an engineered quantum reservoir \cite{PNAS-108-15097,PNAS-110-2693,EPL-88-50003, PRL-112-030602,PRX-7-031044} and consuming additional resources, such as extra information \cite{PRL-123-250606,PRL-124-100603,PRL-106-070401,PRL-120-100601,PRL-118-260603,PRE-96-022108,PRL-120-260601}.

Here we provide another way to exceed the standard Carnot limit by developing a general theory for a microscopic quantum heat engine coupled to two finite heat reservoirs. The finite heat reservoir is characterized by the finite heat capacity which causes the change of the reservoir's temperature in the isothermal process of a standard Carnot cycle. This effect produces a break between the start and the end of a cycle (called internal cycle), and thus we have to introduce an external cycle to compensate the break. As elucidated below, the efficiency of such an engine, with internal and external cycles working collaboratively, could go beyond the Carnot limit over a suitable range of the working substance (WS) without expending any extra quantum resource. In practice, a maximum efficiency of the engine always accompanies an extinct power output due to the adiabatic cyclic process, implying a trade-off between the efficiency and the power output \cite{PRL-98-108301,PRE-81-051129,PRL-111-050601,PRE-90-062124,BJP-46-282,NC-10-202}. Especially, heat engines at the maximum power output are limited by the Curzon-Ahlborn efficiency \cite{AJP-43-22,PRL-95-190602}; Recent results have also shown that the Carnot efficiency can be reached at the finite power output  \cite{PRL-114-050601,EPJB-89-248,EPL-118-40003,PRD-98-026008,PRL-121-120601,NJP-21-103049}. In contrast, we focus on the key role played by the finite heat reservoirs in the present study, and concern the relevance to currently available techniques of laboratory.
Making use of the recent ideas of open quantum systems \cite{PRE-86-011127,PRE-88-062115,PRL-119-050601,PRE-95-012148,PRA-98-012139,PRL-124-110606}, we explore optimal strategies, in an analytical fashion, to reach the maximum efficiency and the maximum power output of the engine surpassing the Carnot limit.

\section{Model and protocol}

\begin{figure}[tbph]
\centering\includegraphics[width=8.0 cm, height=4.0 cm]{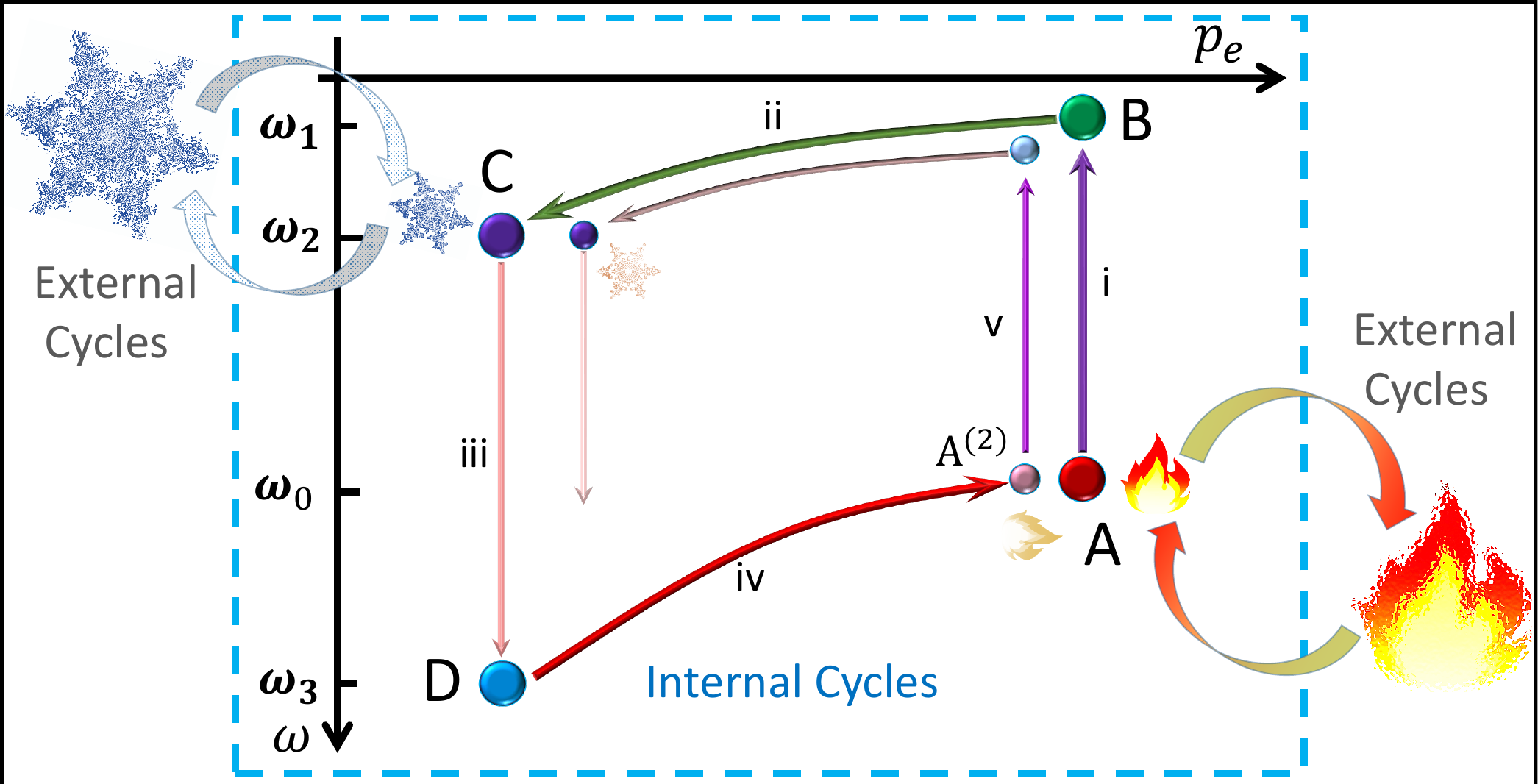}
\caption{
{\bf Schematic for a quantum cyclic heat engine, exemplified by a two-level system as the WS}. Internal cycles: Population-frequency diagram for the internal cycle of a two-level quantum heat engine (elucidated later) with the population of the upper state $p_e=[1-\tanh(\beta_{k}\omega(t)/2)]/2$ in a Gibbs state, where $\beta_{k}$ changes due to contact with the finite heat reservoirs and $\omega(t)$ is the time-varying two-level resonance frequency with $\omega(0)=\omega_0$. In the process from A to B (C to D), the WS executes the isentropic expansion (compression), while the process from B to C (from D to A$^{(2)}$) is an isothermal cooling (heating) by contacting the low-temperature (high-temperature) bath. External cycles: The built-in finite heat reservoirs of the internal cycle are reset to the initial temperature to avoid the engine’s idling by contacting their corresponding infinite cold or heat baths.}
\label{fig1}
\end{figure}

As sketched in Fig. \ref{fig1}, our quantum Carnot cycle consists of two process, i.e., internal cycle process and external cycle process. The internal cycle process consisting of four consecutive steps, i.e., expansion-cooling-compression-heating. The system (i.e., the WS), governed by a time-dependent Hamiltonian, is coupled to two finite heat reservoirs at the initial temperatures $T_h^{(1)}$ and $T_c^{(1)}$ with the superscript (n) implying the $n$th round and $T_h^{(n)}>T_c^{(n)}$ (For simplicity of description, we omit the superscripts below for the first round, i.e., $T_{h,c}^{(1)}=T_{h,c}$). The internal cycle gets started from the starting point $A$ with the WS in thermal equilibrium with the high-temperature reservoir, where the WS is described by the Hamiltonian $H_0$. Each round of the cycle takes following four steps: (i) The system is kept in isolation from the reservoirs, undergoing an adiabatically isentropic expansion process, in which the system's Hamiltonian $H(t)$ varying from $H_0$ to $H_1$ commutes with the initial state at any time $t$. Thus this step is accomplished by a Hamiltonian quench during which the system' state is unaffected. The end of the step (i) is the system in thermal equilibrium with the low-temperature reservoir, implying $H_1=H_0T_c/T_h$. (ii) Contacted with the low-temperature reservoir, the Hamiltonian of the system is isothermally modulated from $H_1$ to $H_2$ with a duration $\tau_c$. In this case, the temperature of the low-temperature reservoir is also changed, which can be labeled as $T_c^{(2)}$, i.e., the beginning temperature of the low-temperature reservoir when interacting with the WS in the second round. (iii) Similar to step (i), a second quench is carried out, followed by a thermal equilibrium with the high-temperature reservoir. Thus, the Hamiltonian of the WS varies from $H_2$ to $H_3=H_2T_h/T_c^{(2)}$ at the end of this step. (iv) Coupled to the high-temperature reservoir again, the system is isothermally back to the Hamiltonian $H_0$ with a duration $\tau_h$, along with the temperature change of the high-temperature reservoir to be $T_h^{(2)}$ at the end of this step. Then the second round of the internal cycle begins, as labeled in Fig. \ref{fig1} as the step (v) and so on. The internal cycle generally needs to go on for several rounds until some termination conditions are satisfied, as clarified later. 

After the internal cycle is accomplished, the periodical reset process, to avoid the engine's idling, is executed by the external cycle of the engine. In this process, we couple these two finite heat reservoires of internal cycles to two baths with infinite capacities, respectively, where the internal low-temperature finite heat reservoir delivers heat into the infinite cold bath with temperature $T_c$ and the internal high-temperature finite heat reservoir absorbs heat from the infinite hot bath with temperature $T_h$. After this process, the internal heat reservoirs will be in equilibrium with the corresponding external heat reservoires.

\begin{figure*}[tbph]
\includegraphics[width=16.0 cm, height=7.5 cm]{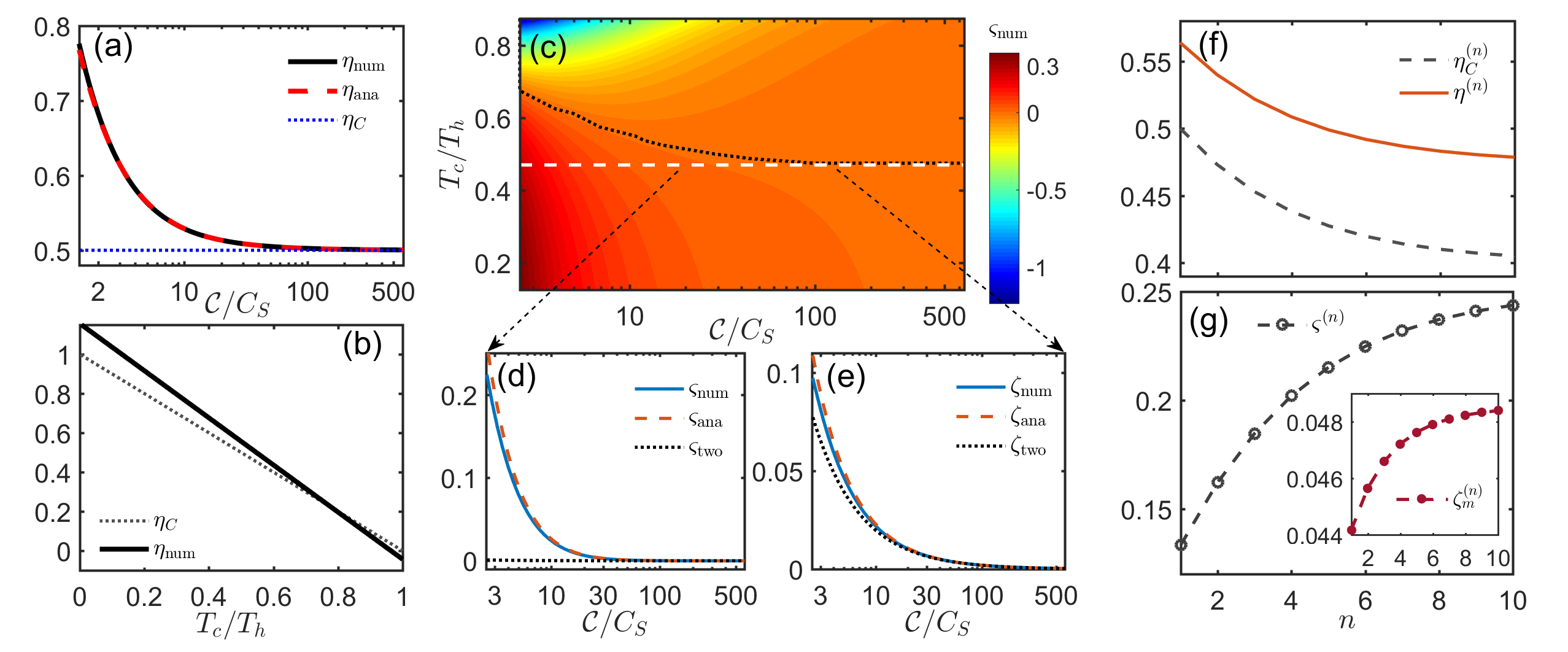}
\caption{Analytical and numerical results of the internal cycle for the two-level system as the WS.  Efficiency of the Carnot engine in the first round with respect to (a) the heat capacity $\mathcal{C}$ with $T_h/T_c=2$ and (b) temperature with $\mathcal{C}/\mathcal{C}_S=6.4$, where the subscripts 'C', 'num' and 'ana' denote, respectively, the Carnot limit, the numerical result and the analytical result in the first round. The efficiency in (b) surpassing the Carnot limit is obtained when $T_c/T_h<2/3$. (c) Enhanced power coefficient $\varsigma$ with respect to temperature ratio and heat capacity, where the dotted curve denotes the zero boundary and the dashed line at $T_c/T_h=3(7-\sqrt{33})/8$ is the zero line obtained from the analytical result of the two-level system. (d) $\varsigma$ and (e) $\zeta$ at the maximum power output along the zero line, where the symbol 'two' denotes the corresponding analytical results of the two-level system given in Discussion. (f) Efficiency of the engine compared with the Carnot limit $\eta_C^{(n)}=1-T_c^{(n)}/T_h^{(n)}$ in the $n$th round. (g) Power output enhancement $\varsigma^{(n)}$ (with its efficiency increment $\zeta^{(n)}$ in Inset) of the $n$th round, where $\varsigma^{(n)}=P_m^{(n)}/P_C^{(n)}-1$, $\zeta^{(n)}=\eta^{(n)}/\eta_{CA}^{(n)}-1$ with the Carnot power output $P_{C}^{(n)}=\Delta S_h^{(n)}(\sqrt{T_h^{(n)}}-\sqrt{T_c^{(n)}})^2/4\iota$ (the Carnot efficiency $\eta_{CA}^{(n)}=1-\sqrt{T_c^{(n)}/T_h^{(n)}}$). Other parameters: $\iota=0.5\omega_0^{-1}$ and $\omega_2T_h/\omega_0T_c=\sqrt{2}$ in (a)-(e) while $1.2$ in (f,g).}
\label{fig2}
\end{figure*}

\section{Internal cycle process}
For our purpose, we first scrutinize the heat exchange in the internal cycle, i.e., in the isothermal processes. For the finite heat reservoirs, we define the heat capacities of the reservoirs as  $\mathcal{C}_k=dU_k/dT_k$ with $k=h,c$ (also in the following) and the internal energy $U_k$ being finite. Considering the first round, the dynamics of the thermodynamic quantities
in the isothermal process (in units of $k_B=\hbar=1$ throughout the paper) can be written as
\begin{equation}
\dot{Q}_k=T_k\dot{S}_k-\dot{W}_k^{ir}, \quad \dot{T}_k=-\dot{Q}_k/\mathcal{C}_k, \notag
\end{equation}
and the work $W_k=\Delta U_K-Q_k$ with $Q_k$, $W_k^{ir}$, $S_k$ and $U_k$ denoting the heat exchange, irreversible work dissipation, entropy and internal energy of the WS, respectively. Then, the heat exchange can be obtained as (see Supplementary Information),
\begin{equation}
Q_k=T_k\Delta S_k-\int_0^{\tau_k}\dot{S}_k\int_0^t\dot{Q}_k\mathcal{C}_k^{-1}d\tau dt- W_k^{ir},
\label{Eq1}
\end{equation}
where $\Delta S_k=S(\pi_k(\tau_k))-S(\pi_k(0))$ with the von Neumann entropy $S(\cdot)$ and the Gibbs state $\pi_k(t)=e^{-\beta_k(t) H_k(t)}/\text{Tr}[e^{-\beta_k(t) H_k(t)}]$ at temperature $\beta_k(t)=1/T_k(t)$. In our case, $\Delta S_c+\Delta S_h\neq 0$ in the internal cycle. The second term comes from the temperature variation of the heat reservoir due to the finite capacity. The irreversible work dissipation in the last term is given by $W_k^{ir}=T_kV_k/\tau_k$ with $V_k$ relevant to the thermodynamic length at finite duration \cite{PRL-124-110606,PRL-51-1127,ZPB-59-449,JCP-83-334,PRL-99-100602,PRL-908-190602,JCP-140-244119,Q-3-197}. Then, the engine's efficiency is given by,
\begin{equation}
\eta=1+\frac{Q_c}{Q_h}+\frac{\text{Tr}[(\pi_c(0)-\pi_h(\tau_h))H_0]}{Q_h},
\label{Eq2}
\end{equation}
where the last term originates from the misalignment of the end of the step (iv) from the starting point of the round. This term is always positive due to the temperature decrease of the high-temperature reservoir, ensuring breakthrough of the Carnot limit.

For a large heat reservoir, i.e., $\mathcal{C}_k\gg |\Delta S_k|$ and keeping the leading terms of $\mathcal{C}_k$, Eq. (\ref{Eq1}) is reduced to
\begin{equation}
Q_k=\tilde{T}_k\Delta S_k(1-\frac{\Delta S_k}{2\mathcal{C}_k}),
\label{Eq3}
\end{equation}
and the temperature $T_k^{(2)}=T_k-\Delta S_k\tilde{T}_k/\mathcal{C}_k$ where the finite duration creates an effective temperature $\tilde{T}_k=(1-\iota_k/\tau_k)T_k$ with the time parameter of the internal cycle $\iota_k:=V_k/S_k$. Furthermore, we can obtain the heat capacity of the WS for Hamiltonian $H_0$ as $\mathcal{C}_S=\text{Tr}[(\pi_c(0)-\pi_h(\tau_h))H_0]/(T_h-T_h^{(2)})$.
Therefore, Eq. (\ref{Eq2}) is reduced to $\eta=\eta_{C}+\zeta$ with the Carnot limit $\eta_{C}=1-T_c/T_h$, and the increment $\zeta$ induced by the finite heat reservoirs, i.e.,
\begin{equation}
\zeta=\frac{\mathcal{C}_S}{\mathcal{C}_h }+\frac{T_c}{T_h}\left[ 1-g_{\tau}\left( 1-\frac{\delta S}{\Delta S_{h}}\right) -\frac{g_{\tau}(\mathcal{C}_c+\mathcal{C}_h)\Delta S_{h}}{2\mathcal{C}_c\mathcal{C}_h}\right],
\label{Eq4}
\end{equation}
where $\delta S=\Delta S_c+\Delta S_h$ with $\delta S\ll \Delta S_{h}$, and the factor caused by the finite duration is defined as $g_{\tau}=\tau_h(\tau_c-\iota_c)/\tau_c(\tau_h-\iota_h)$.

Equation (\ref{Eq4}), showing the potential to surpass the Carnot limit (see Fig. \ref{fig2}(a,b), is one of the main results in the present work. The first term in $\zeta$ represents the efficiency increase due to the finite heat capacities of the high-temperature reservoir and the WS. In contrast, the second term, which is induced by the capacities of the heat reservoirs and the duration of the isothermal process, generally indicates the efficiency decrease. In the large reservoir limit, i.e., $\mathcal{C}_k\rightarrow\infty$, Eq. (\ref{Eq4}) returns to the Carnot efficiency with $\zeta=(1-g_{\tau})T_c/T_h$.
Besides, since the conditions of $\iota_c\leq 0$ and $\iota_h\geq 0$ make $g_{\tau}\geq 1$, the maximal efficiency appears always in the slowly dynamical limit, i.e., $\tau_{c,h}\rightarrow \infty$. However, the engine working at the maximal efficiency will lose its power output due to an infinite duration of the isothermal process. As a result, the key problem here is the maximum power $P=-W/(\tau_c+\tau_h)$ at the finite time, which can be solved by optimizing the durations of the two isothermal processes satisfying the condition $\partial P/\partial\tau_k=0 $. This corresponds to the optimal ratio $\tau_h/\tau_c=\sqrt{T_h\epsilon/\mathcal{C}}$ with the amendment factor $\epsilon$ induced by the finite heat reservoirs (see Supplementary Information). 

For the two reservoirs with the same heat capacity ($\mathcal{C}\equiv\mathcal{C}_h=\mathcal{C}_c$), Eq. (\ref{Eq4}) in the slowly dynamical limit is reduced to $\zeta=(\mathcal{C}_ST_h\Delta S_h-T_c\Delta S_h^2+T_c\mathcal{C}\delta S)/\mathcal{C}T_h\Delta S_h$. Keeping the first-order term of $\mathcal{C}$, we acquire the maximum power output as $P_m=P_C(1+\varsigma)$ in the high-temperature limit, where the maximum power output of the Carnot cycle is $P_C=\Delta S_h(\sqrt{T_h}-\sqrt{T_c})^2/4\iota$ (assuming $\iota:=\iota_h=-\iota_c$) \cite{EPL-81-20003,PRL-124-110606,PRL-105-150603}. The improvement induced by the effect of the finite heat reservoir is given by (see Methods for details)
\begin{equation}
\varsigma=\frac{\sqrt{T_h}\Delta S_h (2\mathcal{C}_S-\Delta S_h)+\sqrt{T_c}(2\mathcal{C}\delta S-\Delta S_h^2)}{2\mathcal{C}(\sqrt{T_h}-\sqrt{T_c})\Delta S_h}.
\label{Eq5}
\end{equation}
Eq. (\ref{Eq5}) is another main result of the present work, which indicates the maximum power output surpassing the power limit of the Carnot cycle if the heat capacities, the entropy change and the temperature ratio $T_c/T_h$ make $\varsigma\geq 0$ (see Fig. \ref{fig2}(c,d)). At the maximum power output, the corresponding efficiency is reduced to $\eta_m=\eta_{CA}+\zeta$ with the Curzon-Ahlborn efficiency $\eta_{CA}=1-\sqrt{T_c/T_h}$ \cite{AJP-43-22} and the nonzero term $\zeta$ caused by the finite heat reservoir 
\begin{equation}
\zeta=\frac{\mathcal{C}_S}{\mathcal{C}}-\frac{1}{2}\sqrt{\frac{T_c}{T_h}}(\frac{\Delta S_{h}}{\mathcal{C}}+\frac{\mathcal{C}_S}{\mathcal{C}}-\frac{\delta S}{\Delta S_{h}}),
\label{Eqb13}
\end{equation}
which is always positive and inversely proportional to the heat capacity of the reservoir (see Fig. \ref{fig2}(e)), implying $\eta_m>\eta_{CA}$.



Since the temperature of the finite heat reservoir varies with the cycling, we have the following iterative relation,
\begin{equation}
T_k^{(n)}=\left[ 1-\frac{\Delta S^{(n-1)}_k}{\mathcal{C}_k}\left( 1-\frac{\iota_k}{\tau_k}\right)\right] T^{(n-1)}_k.
\label{Eq6}
\end{equation}
Despite the shrink of the entropy change as the increase of the internal cycling, Eq. (\ref{Eq6}) shows an exponential variation tendency of  the temperature (see Supplementary Information), which leads to exponential decays of the efficiency and the power output as $n$ increases. As shown in Fig. \ref{fig2}(f,g), both the maximum efficiency and the power output can surpass the corresponding Carnot limit in other rounds except the first. The smaller temperature difference between the two reservoirs favors the surpassing. However, the smaller temperature difference between the two reservoirs would increase the duration of single rounds in the isothermal process, which lowers the power output. Therefore, the temperature difference between the two reservoirs should be set appropriately. Otherwise, the decreasing efficiency and power output of the engine would finally disable the engine with the internal cycle proceeding.

\begin{figure}[t]
\includegraphics[width=9.0 cm, height=4.0 cm]{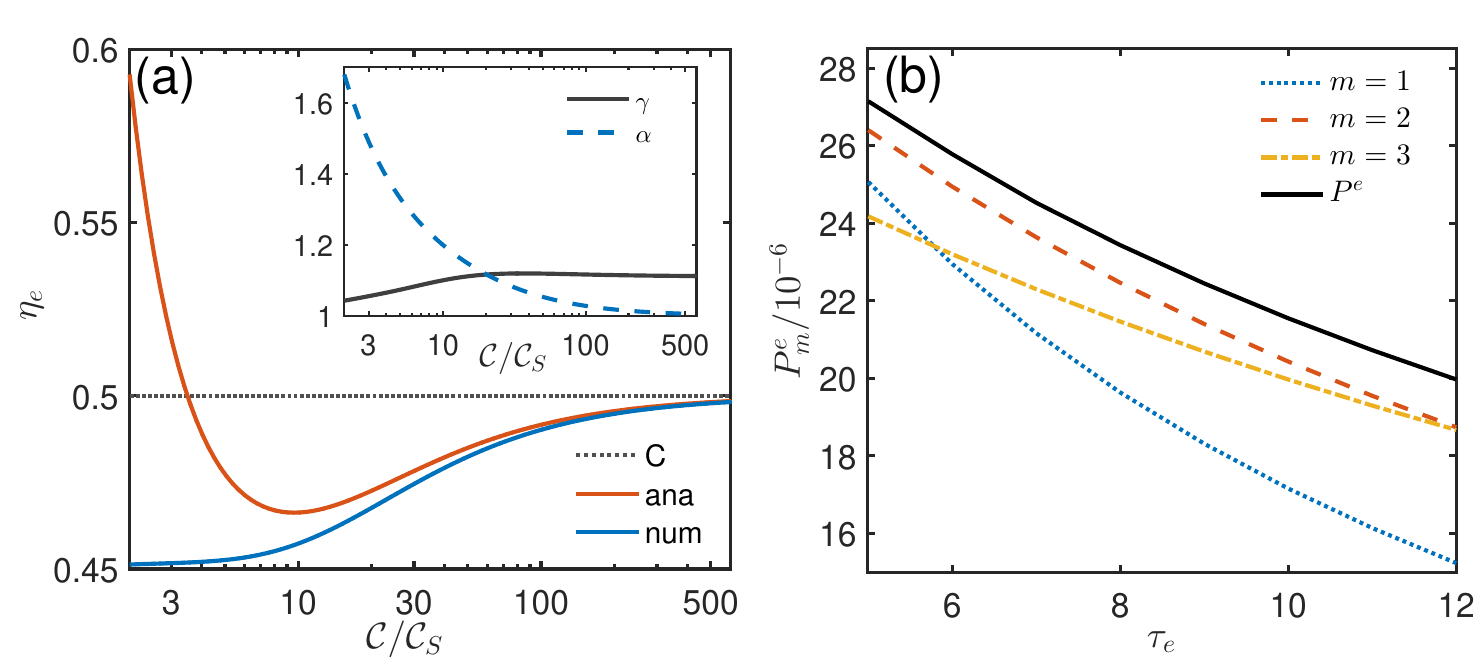}
\caption{ Analytical and numerical results regarding the external cycle and Eq. (8). (a) Efficiency of the external cycle with respect to the heat capacity $\mathcal{C}$ of the reservoirs, where the symbols 'C', 'ana' and 'num' denote the Carnot limit, the analytical result in Eq. (\ref{Eq67}) and the numerical result, respectively. Inset: variation of $\gamma$ and $\alpha$. (b) Average power output of the engine with respect to the reset time, where $m$ denotes the number of rounds taken in the internal cycle and $P^e$ is obtained by calculating Eq. (\ref{Eq7}). The parameters are the same as in Fig. \ref{fig2}(f). }
\label{fig4}
\end{figure}

\section{External cycle process}
In this process, the built-in low-temperature finite heat reservoir delivering heat into an infinite cold bath is cooled down to $T_c$ with the heat exchange of $Q^r_c=\mathcal{C}_c(T_c^{(m+1)}-T_c)$. Meanwhile, an infinite hot bath injects heat into the built-in high-temperature finite heat reservoir and raises the latter's temperature to $T_h$. This also implies the system's temperature reaching $T_h$ due to the fact that the system is coupled to the high-temperature finite heat reservoir at the end of the $m$th round of the internal cycle. Thus, the heat absorption is assessed as $Q^r_h=(\mathcal{C}_h+\gamma\mathcal{C}_S)(T_h-T_h^{(m+1)})$ with $\gamma>0$. In the slowly dynamical limit, we obtain the maximum efficiency of the external cycle as
\begin{equation}
\eta^e=\eta_C+\frac{\alpha\gamma\mathcal{C}_ST_c}{\mathcal{C}T_h} -(\alpha-1)\frac{T_c}{T_h},
\label{Eq67}
\end{equation}
with $\alpha\geq 1$ (see Supplementary Information for details). 
As plotted in Fig. \ref{fig4}(a), the efficiency of the external cycle can approach the Carnot limit in the large reservoir limit, but cannot surpass the Carnot limit \cite{PRE-96-012128,PRE-100-012122} although the analytical result based on some approximations shows the possibility of surpassing for the reservoirs with small heat capacity.

For a finite duration of the isothermal process, the average power output depends on the summed duration of the internal cycle and the reset time. Therefore, obtaining the maximal average power output is a multi-parameter optimization problem in a multi-cycle process, which can be stated as
\begin{equation}
P^e=\max_{m,\tau_k}\frac{\mathcal{C}_S(T_h-T_h^{(m+1)})+\sum^m_{n=1}(Q_c^{(n)}+Q_h^{(n)})}{\sum_{n=1}^m\sum_{k=c,h}\tau_k^{(n)}+\tau_e}.
\label{Eq7}
\end{equation}
In principle, maximizing the power output in each round of the internal cycle could produce a large average power output, even though not the maximal one, of the external cycle. A higher power output always favors a shorter reset time due to no power output during the reset process. Therefore, to acquire a large average power output, we may terminate the internal cycle at the end of the $m$th round for a given reset time $\tau_e$. As demonstrated in Fig. \ref{fig4}(b), for the case of two rounds of the internal cycle, the power output $P^e_2$ is closer to the maximal average power output $P^e$ than others. For $\tau_e<4\tau^{(1)}$, $m=2$ is an optimal solution for the internal cycling due to the trade-off between the cycling number and the reset time $\tau_e$.

\section{Concrete system}

\subsection{Two-level system}
First of all, we exemplify a two-level system as the WS \cite{PRL-123-080602,PRL-122-240602,PRL-123-240601}, such as a qubit encoded in two internal levels of an ion \cite{PRL-120-010601,PRL-120-210601}, described by the Hamiltonian $H(t)=\omega(t)\sigma_z/2$ with $\sigma_z$ being the usual Pauli operator and the energy difference $\omega(t)$ modulated at the two fixed points $\omega_0$ and $\omega_2$ of the internal cycle. This case has been treated as an example in Fig. \ref{fig2} both analytically and numerically. In the first round of the internal cycle, the entropy change is solved as $\Delta S_h=(\omega^2_2T_h^2-\omega^2_0T_c^2)/8T_c^2T_h^2$, guaranteeing an effective power output if $\omega_2/\omega_0>T_c/T_h$. Thus we have the heat capacity of the WS as $\mathcal{C}_S=\omega_0^2/4T_h^2$. In the adiabatic limit, the efficiency for the finite heat reservoirs owning the same heat capacity is enhanced by
$\zeta=[2\omega^2_0T_cT_h-\omega^2_0T_c^2-\omega^2_2T_h^2]/8\mathcal{C}T_cT_h^2(T_h-T_c)$.
Thus the Carnot limit could be surpassed if $ \omega_2/\omega_0<\sqrt{T_c(2T_h-T_c)}/T_h$, see Fig. \ref{fig2}(b). Besides, the coefficient in Eq. (\ref{Eq5}) is now given by
\begin{eqnarray}
\varsigma=\frac{\omega_0^2T_c^2(5T_h-\sqrt{T_cT_h}-2T_c)-\omega_2^2T_h^2(T_h+\sqrt{T_cT_h})}{16\mathcal{C}T_c^2T_h^2(T_h-\sqrt{T_cT_h})},
\label{Eq9}
\end{eqnarray}
which implies the fact going beyond the maximum power output of the Carnot cycle if the temperature ratio $T_c/T_h<r_{\omega}$ for the given modulation frequencies $\omega_{0}$ and $\omega_{2}$. It results in the zero boundary in Fig. \ref{fig2}(c), e.g., $r_{\omega}\approx 0.47$, and vice versa (see Supplementary Information). Meanwhile, the efficiency at the maximum power output will exceed the Curzon-Ahlborn efficiency if $ T_c/T_h<R_{\omega}$ (e.g., $R_{\omega}\approx 0.69$) (see Supplementary Information) that the power output beyond the maximal counterpart of the Carnot limit is always accompanied by its efficiency larger than the Curzon-Ahlborn efficiency, see Fig. \ref{fig2}(e). On the other hand, since the finite heat reservoir should own a larger heat capacity than that of the WS, we consider that an ensemble of spin-$1/2$ particles might be a suitable candidate of the finite heat reservoir \cite{PRL-88-097905,PRA-89-052120,SR-6-33945,NJP-18-123018}. In addition, for the moderate temperature of the WS, i.e., comparable to $\omega_{0,2}$, the harmonic oscillators can be also employed as the finite heat reservoirs in this situation \cite{PRL-123-080602}.

\subsection{Harmonic oscillator system}
Secondly, the harmonic oscillator can also be employed as the WS, such as the quantized vibration of an ion \cite{PRL-112-030602}, described by the Hamiltonian $H(t)=\omega(t)a^{\dagger}a$ with $a$ ($a^{\dagger}$) denoting the annihilation (creation) operator. The entropy change in the first round of the internal cycle is $\Delta S_h=\ln(\omega_2 T_h/\omega_0T_c)$ and the corresponding heat capacity is $\mathcal{C}_S=1$. Thus the maximal efficiency of the engine is enlarged by
$\zeta=\frac{1}{\mathcal{C}}[\eta_C-\frac{T_c}{T_h}\ln\frac{\omega_2T_h}{\omega_0T_c}]$,
which makes $\eta$ beyond the Carnot limit if $\omega_2/\omega_0<T_c\exp(T_h/T_c-1)/T_h$. Besides, the enhancement of the maximal power output is
\begin{eqnarray}
\varsigma=\frac{2\sqrt{T_h}+\sqrt{T_c}}{2\mathcal{C}\sqrt{T_h}}-\frac{\sqrt{T_h}+\sqrt{T_c}}{2\mathcal{C}(\sqrt{T_h}-\sqrt{T_c})}\ln\frac{\omega_2 T_h}{\omega_0T_c}.
\label{Eq11}
\end{eqnarray}
If $\varsigma>0$, we have the restriction of $T_c/T_h<r_{\omega}$ for the given modulation frequencies $\omega_{0,2}$. Besides, the Curzon-Ahlborn efficiency will be exceeded if $T_c/T_h<R_{\omega}$. In practice, a finite multi-mode environment can serve as the heat reservoirs in this case \cite{PRL-85-1799,PRA-90-041108,OE-23-5763}; For the moderate temperature, i.e., comparable to the oscillator's frequency, a spin ensemble can also be employed as the finite heat reservoir.

\section{Discussion and conclusions}

We mention the influence from the possible dissipation, which was neglected above. The influence lies in two aspects. One is from the irreversible work dissipation caused by the finite duration in the cooling and heating process. The other is from the extra heat dissipation of the WS and the reservoirs due to the coupling with the external environment. In the slowly dynamical limit that the duration is infinitely long, the work dissipation caused by the first factor is zero but the second factor would yield the efficiency of the heat engine to be vanishing. In the case of the maximum power output, however, the duration is finite. Hence the main influence is caused by the first factor which restricts the efficiency of the Carnot cycle within the Curzon-Ahlborn (CA) efficiency, whereas the second factor is negligible. Nevertheless, this consideration of dissipation does not change the  main results in our work.

We have explored the possibility going beyond the Carnot limit using a microscopic engine coupled to two finite heat reservoirs.  The favorable feature of our proposal lies in the experimental relevance: No extra quantum resource necessary, no special system required and availability using current laboratory techniques. Our finding paves the way to employing finite-sized reservoirs in microscopic quantum thermodynamics and clarifies the high efficiency of the nano-engines coupled to two same-sized finite heat reservoirs with different temperatures.

\section{Conflict of interest}
The authors declare that they have no conflict of interest.

\section{Acknowledgements}
This work is supported by the National Key Research and Development Program of China under Grant No. 2022YFA1404500, by Cross-disciplinary Innovative Research Group Project of Henan Province under Grant No. 232300421004, National Natural Science Foundation of China under Grant Nos. 1232410, U21A20434, 12074346, 12274376, 12074232, 12125406, 12374466, by Natural Science Foundation of Henan Province under Grant Nos. 232300421075, 242300421212, by Major science and technology project of Henan Province under Grant No. 221100210400, by K. C. Wong Education Foundation (GJTD-2019-15), and by Nansha senior leading talent team technology project under Grant No. 2021CXTD02. 

\section{Author contributions}
Y.Y.L., S.L.S. and M.F. proposed the idea and made the main deductions; Y.Y.L. carried out numerical calculation; Y.Y.L., M.F. and G. Chen wrote the paper; All the authors contributed to the interpretation of results and improvement of the manuscript.

\section{Appendix A. Supplementary materials}
Supplementary materials to this article can be found online.

\end{document}